\documentclass[10pt,twocolumn]{article}

\usepackage{enumitem}
\usepackage{graphicx}
\usepackage{dcolumn}
\usepackage{array,multirow}
\usepackage{bm}
\usepackage{amsmath}
\usepackage{epstopdf}
\usepackage{amsfonts}
\usepackage{amssymb}
\usepackage{epsfig}
\usepackage{tabularx}
\usepackage{color}
\usepackage[colorlinks=true,citecolor=blue,urlcolor=magenta,breaklinks]{hyperref}

\newcommand{\be}{\begin{equation}}
\newcommand{\en}{\end{equation}}
\newcommand{\bea}{\begin{eqnarray}}
\newcommand{\ena}{\end{eqnarray}}

\begin{document}

\begin{titlepage}

%\rightline{hep-th/yymmnnn}

%\vskip 2cm

\centerline{\large \bf {Radial oscillations of boson stars made of ultralight repulsive dark matter}}

\vskip 1cm

\centerline{Il{\'i}dio Lopes and Grigoris Panotopoulos}

\vskip 1cm

\centerline{Centro de Astrof{\'i}sica e Gravita\c c\~ao - CENTRA, Departamento de F{\'i}sica,}

\vskip 0.2cm

\centerline{Instituto Superior T{\'e}cnico-IST, Universidade de Lisboa-UL}

\vskip 0.2 cm

\centerline{Av. Rovisco Pais 1, 1049-001 Lisboa, Portugal}

\vskip 0.5 cm

\centerline{Email:
\href{mailto:ilidio.lopes@tecnico.ulisboa.pt}{\nolinkurl{ilidio.lopes@}tecnico.ulisboa.pt},}
\centerline{Email:
\href{mailto:grigorios.panotopoulos@tecnico.ulisboa.pt}{\nolinkurl{grigorios.panotopoulos@tecnico.ulisboa.pt}}}

\begin{abstract}
We compute the lowest frequency radial oscillation modes of boson stars. It is assumed that the object is made of pseudo-Goldstone bosons subjected to a scalar potential that leads to a repulsive self-interaction force, and which is characterized by two unknown mass scales $m$ (mass of the particle) and $F$ (decay constant). First we integrate the Tolman-Oppenheimer-Volkoff equations for the hydrostatic equilibrium of the star, and then we solve the Sturm-Liouville boundary value problem for the perturbations using the shooting method. The effective potential that enters into the Schr{\"o}dinger-like equation as well as several associated eigenfunctions are shown as well. Moreover, we found that the large frequency separation, i.e. the difference between consecutive modes, is proportional to the square root of the mass of the star and the cube of the mass scale defined by $\Lambda \equiv \sqrt{m F}$.
\end{abstract}

%\keywords{3D gravity; Black holes; Hawking radiation; Relativistic scattering theory.}

\end{titlepage}

\section{Introduction}

Since the pioneer work of F.~Zwicky about the dynamics of the Coma galaxy cluster in the 30's \cite{zwicky}, and the observations made by V.~Rubin to determine the rotation curves of galaxies a few decades later \cite{rubin}, we are convinced that most of the non-relativistic matter in the Universe is dark, usually referred to as cold dark matter. In modern times current well-established data coming from many different sources confirm the existence of dark matter \cite{turner}, although its nature and origin still remains a mystery. For a review on dark matter see \cite{DM1,DM2}, and for recent reviews on dark matter detection searches see \cite{DM3,DM4,DM5}. 

\smallskip

Usually dark matter in the standard parametrization of the Big-Bang cosmological model is assumed to be made of weakly interacting massive particles, a conjecture which works very well at large (cosmological) scales ($\ge Mpc$), but encounters several problems at smaller (galactic) scales, like the core-cusp problem, the diversity problem, the missing satellites problem and the too-big-to-fail problem \cite{2017arXiv170502358T}. These problems may be tackled in the context of self-interacting dark matter \cite{spergel1,spergel2}, as any cuspy feature will be smoothed out by the 
dark matter collisions. In addition, if dark matter consists of ultralight scalar particles with a mass $m \leq eV$, and with a small repulsive quartic self-interaction a Bose-Einstein condensate (BEC) may be formed with a long range correlation. This scenario has been proposed as a possible solution to the aforementioned problems at galactic scales \cite{proposal1,proposal2,proposal3}. 

\smallskip

An excellent dark matter candidate with self-interactions is the Quantum Chromodynamics (QCD) axion \cite{axion1,axion2,axion3,axion4,axion5}, a hypothetical spin 0 particle proposed by the Peccei-Quinn mechanism \cite{PQ1,PQ2} to solve dynamically the so-called strong CP problem \cite{2016PhR...643....1M}, which can be stated as "why is the $\theta$ parameter in QCD so small?". The QCD axion, well studied in the literature, is one of the good dark matter candidates \cite{taoso}, although it has not been detected yet. What is more, within String Theory \cite{ST1,ST2}, other ultralight axions with a wide range of masses, have been shown to be good dark matter candidates \cite{string1,string2,string3}.

\smallskip

The QCD axion is a member of a more general class of particles, called pseudo-Goldstone bosons, which are very common in Particle Physics, and they arise in the following way: An additional global symmetry is introduced, which is spontaneously broken at some high scale $F$ with soft explicit symmetry breaking at a lower scale $\Lambda \ll F$. Since the additional symmetry is a spontaneously broken global symmetry, there must be a Goldstone boson associated with this symmetry. However, due to the soft explicit symmetry breaking, this boson acquires a small non-vanishing mass, given by $m=\Lambda^2/F$, instead of being massless. QCD axions have been used in Cosmology to accelerate the Universe, either in a concrete inflationary scenario called Natural Inflation \cite{NI1,NI2,NI3,NI4}, or to describe the current cosmic acceleration \cite{ADE1,ADE2}. Its self-interaction, however, is found to be attractive, while tackling the cusp/core problem, as already mentioned before, requires a repulsive force.

\smallskip

Axion stars are gravitationally bounded collections of axions. The possibility of the formation of axion stars was considered for the first time in \cite{tkachev}. Since then axion stars have been studied in \cite{Ref1,Ref2,Ref3,BMZ,chavanis3,freese}. Boson stars have been studied in \cite{BS1,BS2,BS3,BS4,Harko,Mielke2,BS5,BS6,BS7,BS8}. See also \cite{chavanis1,chavanis2} for Newtonian self-gravitating Bose-Einstein condensates. The maximum mass for bosons stars in non-interacting systems was found in \cite{BS1,BS2}, while in \cite{BS3,BS4} it was pointed out that self-interactions can cause significant changes. In \cite{BS5,BS6} the authors constrained the boson star parameter space using data from galaxy and galaxy cluster sizes. 

\smallskip

It is well-known that the properties of stars, such mass and radius, depend crucially on the equation-of-state, which unfortunately is poorly known. Given the recent advances in Helioseismology and Asteroseismology in general, studying the oscillations of stars and computing the frequency modes offer us the opportunity to probe the interior of the stars and learn more about the equation-of-state, since the precise values of the frequency modes are very sensitive to the thermodynamics of the internal structure of the star \cite{2012RAA....12.1107T}. In this work we develop an identical strategy to study the internal properties of these hypothetical axions stars which are on the list of a future astronomical observational programs \cite{2017ApJ...845L...4K}. For previous works on radial oscillations of stars see \cite{Cox,Frandsen, Aerts,Hekker,chanmugan2,kokkotas,hybrid,basic,LP} and references therein.

\smallskip

It is the goal of the present work to study radial oscillations of Dark BEC stars made of ultralight repulsive scalar particles identified as pseudo-Goldstone bosons. Our work is organized as follows: after this introduction, we present the equation-of-state in section two, and the equations for the hydrostatic equilibrium in the third section. The equations for the perturbations are discussed in section four, and our numerical results in the fifth section. Finally we conclude in the last section. We work in natural units in which $c=\hbar=1$. In these units all dimensionfull quantities are measured in GeV, and we make use of the conversion rules $1 m = 5.068 \times 10^{15} GeV^{-1}$, $1 kg = 5.610 \times 10^{26} GeV$ and $1 sec = 1.519 \times 10^{24} GeV^{-1}$ \cite{guth}.

\section{Hydrostatic equilibrium}

\subsection{Equation-of-state}

The perturbative Lagrangian of a relativistic real scalar field $\phi$ is given by 
\begin{equation}
\mathcal{L} = \frac{1}{2} \partial_\mu \phi \partial^\mu \phi - V(\phi)
\end{equation}
where the scalar potential is of the form
\begin{equation}
V(\phi) = \frac{1}{2} m^2 \phi^2 + \frac{1}{24} \: \frac{m^2}{F^2} \phi^4  + ...
\end{equation}
and where we consider renormalizable theories only, ignoring all higher order terms. In this work the scalar field is identified with any pseudo-Goldstone boson. The sign of the quartic self-interaction is taken to be positive since we assume a repulsive self-interaction for the Dark pseudo-Goldstone boson. Therefore, the model assumed here is characterized by two unknown mass scales, namely the mass of the scalar particle, $m$, as well as the decay constant, $F \gg m$, arising from the spontaneous breaking of some global symmetry. Unfortunately, it turns out that it is not easy to obtain scalar field models with a tiny mass and a repulsive force within known Particle Physics, although some attempts have been made \cite{Fan}. In the following, without relying on concrete Particle Physics models, we shall assume that this is possible, and we shall study radial oscillations of objects made of Dark pseudo-Goldstone bosons.

\smallskip

The above scalar potential combined with the Gross-Pitaevskii equation \cite{BEC1,BEC2,BEC3}, also known as non-linear Schr{\"o}dinger equation, leads to the following equation-of-state for the ultralight 
pseudo-Goldstone boson \cite{chavanis3,Fan}:
\begin{equation}
P(\epsilon) = K \epsilon^2
\end{equation}
where the constant $K$ is computed to be \cite{chavanis3,Fan}
\begin{equation}
K = \frac{1}{(2 \Lambda)^4}
\end{equation}
where a new mass scale $\Lambda \equiv \sqrt{m F}$ has been introduced.

\smallskip

The interested reader can find more details on the properties of the Bose-Einstein condensate in the appendix in the end of the article.

\subsection{Structure equations}

We briefly review relativistic stars in General Relativity. The starting point is Einstein's field equations without a cosmological constant
\be
G_{\mu \nu} = R_{\mu \nu}-\frac{1}{2} R g_{\mu \nu}  = 8 \pi G_N T_{\mu \nu}
\en
with $G_N$ being Newton's constant, which is the following will be set to unity. In the exterior of the star the matter energy momentum
tensor $T_{\mu \nu}$ vanishes. For matter we assume a perfect fluid with pressure $P$, energy density $\epsilon$ and an equation of state
$P(\epsilon)$.
For the metric in the case of static spherically symmetric spacetimes, we consider the following ansatz
\be
ds^2 = -e^{A(r)} dt^2 + e^{\lambda(r)} dr^2 + r^2 (d \theta^2 + sin^2 \theta d \varphi^2),
\en
with two unknown metric functions of the radial distance $A(r)$ and $\lambda(r)$.
For the exterior problem $r > R$, with $R$ being the  radius of the star, one obtains the well-known Schwarzschild solution
\be 
 e^{A(r)}= e^{-\lambda(r)} = 1-\frac{2 M}{r},
\en
where $M$ is the mass of the star. For the interior solution $r < R$, we introduce the mass function, $\lambda(r)=-\ln{(1-2m(r)/r})$. The continuity of the model is guaranteed by  matching the two solutions at the surface of the star, for which we obtain $m(R)=M$.

\smallskip 
 
The Tolman-Oppenheimer-Volkoff (TOV) equations for the interior solution \cite{OV,tolman},  with a vanishing cosmological constant read
\bea
m'(r) & = & 4 \pi r^2 \epsilon(r), 
\ena
\bea
P'(r) & = & - (P(r)+\epsilon(r)) \: \frac{m(r)+4 \pi P(r) r^3}{r^2 (1-\frac{2 m(r)}{r})} 
\ena
and
\bea
A'(r) & = & -2 \frac{P'(r)}{\epsilon(r)+P(r)},
\ena
where the prime denotes differentiation with respect to $r$
(i.e., $m^{\prime}\equiv dm/dr$). The first two equations are to be integrated with the initial conditions
$m(r=0)=0$ and $P(r=0)=P_c$, where $P_c$ is
the central pressure. The radius of the star is determined requiring that the energy density vanishes at the surface,
$P(R) = 0$, and the mass of the star is then given by $M=m(R)$. Finally, the other metric function can be computed using the third equation 
together with the boundary condition $A(R)=\ln{(1-2M/R)}$.

\smallskip

To avoid confusion a couple of remarks are in order here. First, since the wavefunction of the Dark BEC star extends to infinity, the radius of the star is formally infinite. This is the reason why in the literature it is used the symbol $R_{99}$ (it contains the $99~\%$ of the mass of the star) rather that $R$. In the following for simplicity we drop the subindex 99. Furthermore, in \cite{Mielke1} in solving the Klein-Gordon equation coupled to Einstein's field equations for any scalar field potential the authors noticed that the boson star was not isotropic, since the corresponding radial pressure component $p_r$ and the tangential pressure component $p_t$ were different, $p_r \neq p_t$, and so there was a non-vanishing anisotropy $p_t-p_r$. In more recent works, however, bosonic dark matter inside the star was modelled as a Bose-Einstein condensate characterized by a polytropic EoS of the form $p=K \rho^2$, and the standard TOV equations for isotropic fluids were considered \cite{DS1,DS2}. This is possible within the Thomas-Fermi approximation where the fractional pressure anisotropy is small, see e.g. \cite{Harko} for its application to self-gravitating bosons stars.

\section{Radial oscillations}

\subsection{Equations for the perturbations}

If $\delta r$ is the radial displacement and $\delta P$ is the   perturbation of the pressure, the equations governing the dimensionless
quantities $\xi=\delta r/r$ and $\eta=\delta P/P$ are the following \cite{chanmugan2,LP,chanmugan1}
\be
\xi'(r) = -\frac{1}{r} \left( 3 \xi + \frac{\eta}{\gamma} \right) - \frac{P'(r)}{P+\epsilon} \xi(r),
\en
and
\begin{equation*}
\begin{split}
	\eta'(r) = \xi \left[ \omega^2 r (1+\epsilon/P) e^{\lambda-A} - \frac{4 P'(r)}{P}  \right]\\
	-\xi \left[ 8 \pi (P+\epsilon) r e^{\lambda} - \frac{r (P'(r))^2}{P (P+\epsilon)}\right] \\
	+ \eta \left[ -\frac{\epsilon P'(r)}{P (P+\epsilon)}-4 \pi (P+\epsilon) r e^{\lambda} \right],
\end{split}
\end{equation*}
where $e^{\lambda}=(1-2m/r)^{-1}$ and $e^A$ are the two metric functions, $\omega$ is the circular frequency oscillation mode
(with  $\nu=\omega/(2\pi)$ being the ordinary frequency), and $\gamma$ is the relativistic  adiabatic index defined to be 
\begin{equation}
\gamma = \frac{d P}{d \epsilon} (1+\epsilon/P).
\end{equation}
Here, however, we prefer to work equivalently with a second order differential equation used in \cite{basic}
\begin{equation}
-f^2 \zeta'' + G \zeta' + (H-\omega^2) \zeta = 0,
\end{equation}
supplemented with the boundary conditions at the origin $r=0$ and at the surface of the star $r=R$: $\zeta(r=0) = 0$ and $\delta p(r=R) = 0$. In the previous equation $\zeta=r^2 e^{-A/2} \xi$ and 
the background functions $f$,  $G$ an $H$  are given by
\begin{equation}
f^2(r) =  \frac{\gamma P e^{A-\lambda}}{P+\epsilon}, 
\end{equation}

\begin{equation}
G(r) =   -\frac{f^2}{\gamma P}
\left[ \frac{\gamma P}{2} (\lambda+3A) + (\gamma P)' - \frac{2 \gamma P}{r} \right], 
\end{equation}

\begin{equation}
H(r) = - \frac{f^2}{\gamma P} \left[ \frac{4 P'}{r} + 8 \pi P (P+\epsilon) e^\lambda - \frac{(P')^2}{P+\epsilon} \right], 
\end{equation}
and finally  the  perturbation of the pressure 
can be computed as 
\begin{equation}
\delta p(r) = -\frac{e^{A/2}}{r^2} ( \zeta P' + \gamma P \zeta' ).
\end{equation}
Therefore, contrary to the previous hydrostatic equilibrium problem, which is an initial value problem, this is a Sturm-Liouville boundary value problem, 
and as such the frequency $\nu$ is allowed to take only particular values, the so-called eigenfrequencies $\nu_n$. Each $\nu_n$ corresponds to a specific radial oscillation mode of the star. Accordingly, each radial mode is identified by its $\nu_n$ and by an associated eigenfunction $\zeta_n$.

\smallskip

The Sturm-Liouville boundary value problem at hand can be treated equivalently as a quantum mechanical problem by recasting the second order differential equation for $\zeta$ into a Schr{\"o}dinger-like equation \cite{2001MNRAS.321..615L} of the form
\begin{equation}
\frac{d^2 \psi}{d \tau^2} + \left[ \omega^2 - U(\tau) \right] \psi = 0,
\end{equation}
where the new variables $\tau$ and $ \psi$ are defined as acoustic radius
$ \tau  =  \int_0^r f^{-1}(z) dz$ and  $ \psi (\tau)  = \zeta/u$.
The effective potential is found to be
\begin{equation}
U = H + \frac{\Pi^2}{4}+\frac{f \Pi'}{2},
\end{equation}
where the function $\Pi$ is given by
$ \Pi  =  -f'-G/f $
while $u$ is determined by the condition
$ u'/u= -\Pi/(2f)$.

\begin{table}
\caption{Boson star's lowest radial oscillation modes 
(for $\Lambda=330~MeV$) for $M=1.86 \times 10^{-14}~M_{\odot}$.}	
\begin{tabular}{l | l}
radial & frequency \\
order & $\nu_n$ \\
$n$ & $(mHz)$ \\
\hline
\hline
		0  & 0.8457 \\
		1  & 1.7426 \\
		2  & 2.5456 \\
		3  & 3.3155 \\
		4  & 4.0676 \\
		5  & 4.8110 \\
		6  & 5.5579 \\
		7  & 6.3032 \\
		8  & 7.0359 \\
		9  & 7.7774 \\
		10 & 8.5076
\end{tabular}
\centering
\label{table:Firstset}
\end{table}

\begin{table}
\caption{Boson star's lowest radial oscillation modes 
(for $\Lambda=330~MeV$) for $M=9.3 \times 10^{-10}~M_{\odot}$.}	
\begin{tabular}{l | l}
radial & frequency \\
order & $\nu_n$ \\
$n$ & $(Hz)$ \\
\hline
\hline
		0  & 0.19 \\
		1  & 0.38 \\
		2  & 0.57 \\
		3  & 0.75 \\
		4  & 0.92 \\
		5  & 1.10 \\
		6  & 1.28
\end{tabular}
\centering
\label{table:Secondset}
\end{table}

\subsection{Numerical results}

We consider for the ultralight scalar boson a mass $m=10^{-8}~eV$, and a decay constant of the order of the GUT scale, $F \sim 10^{16}~GeV$, which correspond to a mass scale $\Lambda \equiv \sqrt{m F}=330~MeV$, and central pressures that correspond to a boson star with a mass $M_1=1.86 \times 10^{-14}~M_{\odot}$ (light object) and $M_2=9.3 \times 10^{-10}~M_{\odot}$ (heavy object) and radius $R=6.9~km$. Therefore, the order of magnitude of the fundamental mode is expected to be \cite{textbook} $\omega_0 \propto \sqrt{M/R^3} = 2.74~mHz$ for the light star and $\omega_0 \propto 0.62~Hz$ for the heavy one, while the spacing in frequencies $\nu_n=\omega_n/2 \pi$ for the highly excited modes is computed by \cite{1994A&A...290..845L,ilidio}
\begin{equation}
\nu_0 = \left[ 2 \int_0^R \frac{dr}{c_s(r)} \right]^{-1},
\end{equation}
where $c_s$ is the sound speed given by $c_s^2 = dP/d \epsilon$.
Conveniently, $\nu_0 $ reads  
\begin{equation}
\nu_0 = \left( \frac{\pi}{2K_2^3} \right)^{1/4} \frac{\sqrt{M}}{a_o},
\end{equation}
where $a_o$ is a numerical constant, such that  $a_o=\int_0^\pi \sqrt{z/sin{z}}\,  dz \approx 5.8993$. In particular, we notice that $\nu_0$ is proportional to $\Lambda^3$ and it scales as $\sim M^{1/2}$. For a given mass one can easily compute the constant spacing at higher excited modes, and for the masses considered here we find $\nu_0 = 0.73~mHz$ for the light star and $\nu_0 = 0.16~Hz$ for the heavy star.

\smallskip

\begin{figure}[ht!]
\centering
{\hspace{-1.0cm}
\includegraphics[scale=0.60]{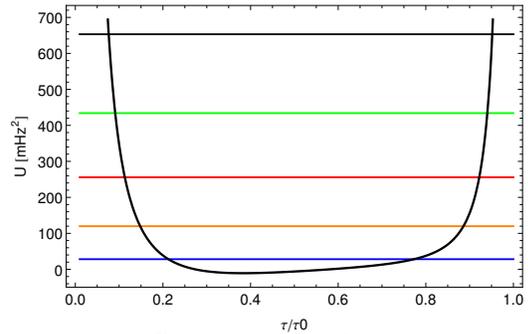}
\vspace{-0.5cm}
}
\caption{Effective potential $U$ (in $(mHz)^2$) versus dimensionless time $\tau/\tau_0$, and the first five radial oscillation frequencies for $\Lambda=330~MeV$ and for $M=1.86 \times 10^{-14}~M_{\odot}$.
The horizontal lines give the frequencies of a few radial modes (n=0,1,2,3,4).}
\label{fig:potential1} 	
\end{figure}

\begin{figure}[ht!]
\centering
{\hspace{-1.0cm}
\includegraphics[scale=0.75]{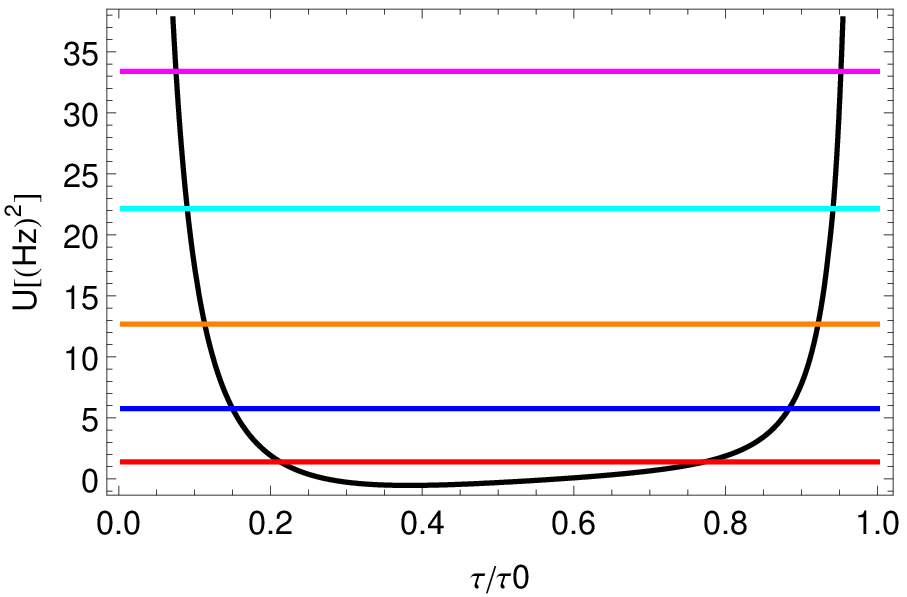}
\vspace{-0.5cm}
}
\caption{Effective potential $U$ (in $(Hz)^2$) versus dimensionless time $\tau/\tau_0$, and the first five oscillation modes for $\Lambda=330~MeV$
and $M=9.3 \times 10^{-10}~M_{\odot}$.
The horizontal lines give the frequencies of a few radial modes (n=0,1,2,3,4).}
\label{fig:potential2} 	
\end{figure}

\begin{figure}[ht!]
\centering
{\hspace{-1.0cm}
	\includegraphics[scale=0.70]{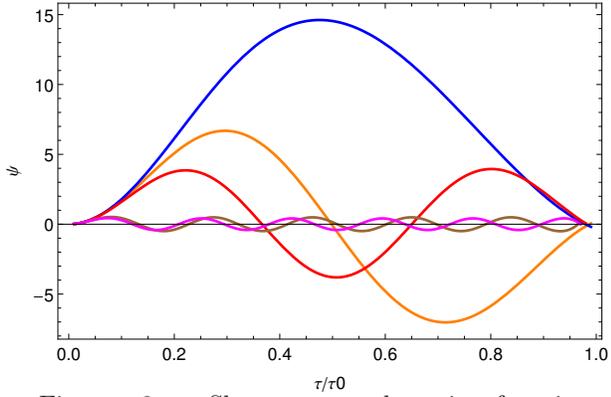}
	\vspace{-0.5cm}
}
\caption{Shown are the eigenfunctions $\psi$ (n=0,1,2,9,10) versus dimensionless time $\tau/\tau_0$ for the fundamental mode (blue), the first and second excited modes (orange and red respectively), and the 9th and 10th excited modes (brown and magenta respectively) for $\Lambda=330~MeV$ and $M=1.86 \times 10^{-14}~M_{\odot}$.}
\label{fig:functions1} 	
\end{figure}

\begin{figure}[ht!]
\centering
{\hspace{-1.0cm}
	\includegraphics[scale=0.9]{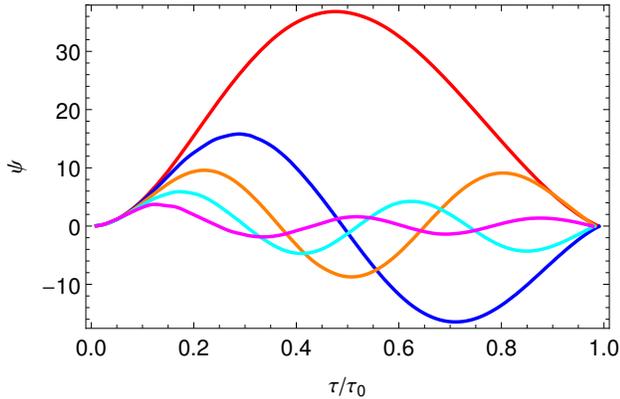}
	\vspace{-0.5cm}
}
\caption{Shown are the eigenfunctions $\psi$ (n=0,1,2,3,4) versus dimensionless time $\tau/\tau_0$ for the fundamental mode as well as the first four excited modes for $\Lambda=330~MeV$ and $M=9.3 \times 10^{-10}~M_{\odot}$.}
\label{fig:functions2} 	
\end{figure}

We computed the lowest oscillation modes summarized in tables~\ref{table:Firstset} and ~\ref{table:Secondset}. We see that the fundamental mode is of the expected order of magnitude. Introducing $\tau_0=\tau(R)=11.4~min$ for the light object and $\tau_0=3.07~sec$ for the heavy one, the effective potential $U$ that enters into the Shr{\"o}dinger-like equation versus dimensionless $\tau/\tau_0$ is shown in Figures~\ref{fig:potential1} and~\ref{fig:potential2} for the light and the heavy star, respectively, while the corresponding eigenfunctions of several modes are shown in Figures~\ref{fig:functions1} and ~\ref{fig:functions2}. We recall that in a Sturm-Liouville boundary value problem the number of zeros of the eigenfunctions corresponds to the overtone number $n$, namely the first excited mode corresponds to $n=1$ and has only one zero, the second excited mode corresponds to $n=2$ and has two zeros, while the fundamental mode does not have zeros and corresponds to $n=0$. In Fig.~\ref{fig:functions1} we show the eigenfunctions $\psi$ corresponding to the fundamental mode, $n=0$ (blue), the first two excited modes, $n=1$ (orange), and $n=2$ (red), and the last two (9th excited:brown and 10th excited:magenta). In Fig.~\ref{fig:functions2} we show the eigenfunctions corresponding to the fundamental and to the first four excited modes.

\begin{figure}[ht!]
\centering
{\hspace{-0.2cm}
	\includegraphics[scale=0.75]{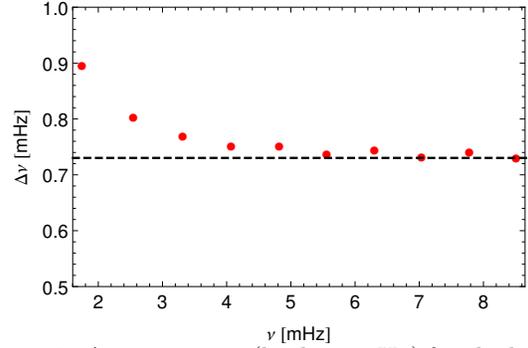}
	\vspace{-0.5cm}}
\caption{$\Delta \nu_n$ versus $\nu_n$ (both in $mHz$) for the lowest acoustic modes for $\Lambda=330~MeV$ and $M=1.86 \times 10^{-14}~M_{\odot}$. The horizontal line corresponds to the constant spacing $\nu_0=0.73~mHz$.}
\label{fig:spectrum1} 	
\end{figure}

\begin{figure}[ht!]
\centering
{\hspace{-0.2cm}
	\includegraphics[scale=0.75]{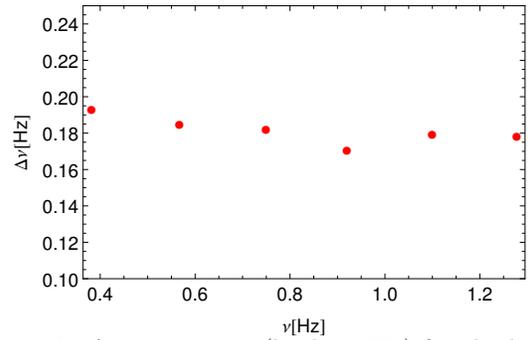}
	\vspace{-0.5cm}}
\caption{$\Delta \nu_n$ versus $\nu_n$ (both in $Hz$) for the lowest acoustic modes for $\Lambda=330~MeV$ and $M=9.3 \times 10^{-10}~M_{\odot}$.}
\label{fig:spectrum2} 	
\end{figure}

Finally, in Figures~\ref{fig:spectrum1} and~\ref{fig:spectrum2} we show the large frequency separation $\Delta \nu_n = \nu_{n+1}-\nu_n$ versus frequencies themselves for $\Lambda=330~MeV$ for the light and the heavy star, respectively. At higher order $n$ the spectrum exhibits a series of equally spaced modes, precisely as expected.

\smallskip
 
Before we finish a couple of remarks are in order about radial and non-radial oscillations of the boson stars considered here. First, regarding detection and oscillations of Dark BEC stars, when these objects collide with neutron stars, or with other objects with a strong magnetic field, depending on how ordinary matter interacts with the ultralight pseudo-Goldstone bosons, fast radio bursts may be emitted in the surface of the neutron star. Such a possibility has been pointed out in the literature for axion stars \cite{Raby}. In such collisions, the tidal forces of the neutron star may produce large variations on the structure of these light Dark BEC stars (that stretch in the direction of the neutron star), which will excite its radial and non-radial oscillations modes \cite{Raby}. These perturbations in the boson stars will then lead to significant variations in their electromagnetic radiation. From the event rate in a galaxy we can determine the mass of the boson star \cite{radio1}. Furthermore, current and future radio telescopes on Earth such as Arecibo or SKA, and the next generation of stellar missions such as "PLAnetary Transits and Oscillation of stars" (PLATO) \cite{plato}, have the potential to put important constraints in their properties, or find such stars if they actually exist. 
 
\smallskip

On the other hand, it would be useful if there were other complementary ways to test the existence of the boson stars and probe their interior. To this end, the Love number formalism can be of great help to characterize the deformation of the star due to the gravitational forces of the companion object. In binary systems the tidal forces may have observable effects on the emission of electromagnetic and gravitational radiation. In particular, the quadrupolar deformation of the star, $Q_{ij}$, is proportional to the external tidal tensor $E_{ij}$ \cite{tidal,maselli}
\begin{equation}
Q_{ij} = - \frac{2 R^5}{3} \: k_2 \: E_{ij} \equiv - \lambda E_{ij} 
\end{equation}
where $\lambda$ is the tidal de-formability, and $k_2$ is the Love number. These two parameters depend entirely on the EoS of the star, and the Love formalism has been proven to be a powerful tool to discriminate between regular black holes, for which $k_2=0$, and other alternatives, such as Exotic Compact Objects \cite{pani}.
In the Newtonian limit, $P \ll \epsilon, \beta \rightarrow 0$, with $\beta=M/R$ being the compactness of the star, and for a polytropic power $\gamma=2$ (polytropic index unity) the Love number is computed to be $k_2 = (\pi^2-5)/(2 \pi^2) \simeq 0.26$ \cite{tidal}, which agrees with the tabulated values of \cite{Hinderer:2007mb}. This value for a BEC Dark star (more general for polytropes with $n=1$) is larger than the highest one for neutron stars with realistic equations-of-state, $\sim 0.14$ \cite{tidal}.

\smallskip

Moreover, it has been proposed that very light dark matter clumps may be detected by Pulsar Timing Arrays \cite{kkmo,AK}, while space-based detectors, such as eLISA, will probe gravitational waves (GWs) in the mHz range \cite{Amaro-SeoaneP,IPL15}. The amplitude of the GW from such light objects is expected to be extremely low. If, however, a huge number of boson stars pulsates simultaneously, the amplitude of the stochastic gravitational wave background may be significantly enhanced.

\section{Conclusions}

We have studied radial oscillations of Dark BEC stars made of ultralight repulsive scalar particles. Although it is highly non-trivial to obtain a scalar potential with a tiny mass and a repulsive self-interaction, we have assumed in this work that this is possible without relying on concrete Particle Physics models. The model is characterized by two free, unknown mass scales, namely the mass of the scalar boson and the decay constant arising from the spontaneous breaking of a global symmetry. First we solved the Tolman-Oppenheimer-Volkoff equations for the hydrostatic equilibrium of the star, and then using the known background solutions we solved the Sturm-Liouville boundary value problem for the perturbation with the shooting method. We have computed the fundamental as well as several excited modes for two different star masses, and we have shown graphically i) several eigenfunctions corresponding to the first three and two highly excited oscillation modes, and ii) how the large frequency difference varies with the frequencies themselves. In addition, we have reformulated the boundary value problem equivalently by writing down a Schr{\"o}dinder-like equation, and we have shown the effective potential together with the first five values of $\omega^2$.

%%%%%%%%%%%%%%%%%%%%%%%%%%%%%%%%%%%%%%%%%%%%%%%%%%%%%%%%%%%%%%%%%%%%%%%%%%%%%%%%%%%%%%

\section*{Acknowlegements}

We are grateful to the anonymous reviewer for a careful reading of the manuscript, for his/her constructive criticism, and useful comments and suggestions. It is a pleasure to thank K.~Clough, D.~Hilditch and C.~Moore for enlightening discussions. The authors thank the Funda\c c\~ao para a Ci\^encia e Tecnologia (FCT), Portugal, for the financial support to the Center for Astrophysics and Gravitation-CENTRA,  Instituto Superior T\'ecnico,  Universidade de Lisboa, through the Grant No. UID/FIS/00099/2013.

%%%%%%%%%%%%%%%%%%%%%%%%%%%%%%%%%%%%%%%%%%%%%%%%%%%%%%%%%%%%%%%%%%%%%%%%%%%%%%%%%%%%%%

\appendix

\section{A cold dilute boson gas}

We model the collection of ultralight scalar particles in the Dark BEC star as a cold dilute boson gas. Under these conditions only binary collisions at low energy are relevant, and thus they are characterized by the s-wave scattering length $l$ irrespectively of the details of the two-body potential. Therefore, we can replace the full potential by a short range repulsive delta-potential of the form \cite{BEC1}
\begin{equation}
{\cal U}(\vec{r}_1-\vec{r}_2) = \left( \frac{4 \pi l}{m} \right) \delta^{(3)}(\vec{r}_1-\vec{r}_2),
\end{equation}
which implies a self-interaction cross section given by $\sigma = 4 \pi l^2$. The ground state properties of the boson gas are described by the mean-field Gross-Pitaevskii equation, also known as non-linear Schr{\"o}dinger equation
\be
i \frac{\partial \Psi(t,\vec{r})}{\partial t} = \left[-\frac{\nabla^2}{2m} + m \: {\cal U} + {\cal U}_0 |\Psi(t,\vec{r})|^2 \right] \Psi(t,\vec{r}),
\en
where ${\cal U}_0=4 \pi l/m$. Using the Madelung representation of the wave function \cite{BEC1,BEC2,BEC3} $\Psi = \sqrt{\rho} \: e^{i S}$ (polar representation of a complex number, with $S$ being the phase and $|\Psi|=\sqrt{\rho}$ being the amplitude), we obtain the following system of two coupled equations for the density $\rho$ and the velocity $\vec{v}=\nabla S/m$ of the quantum fluid
\begin{equation}
\frac{\partial \rho}{\partial t} + \nabla . (\rho \vec{v}) = 0,
\end{equation}
and 
\begin{equation}
\rho \left[ \frac{\partial \vec{v}}{\partial t} + (\vec{v} . \nabla) \vec{v} \right] = -\nabla P -\nabla V_Q - \rho \nabla (V/m),
\end{equation}
where we have introduced the quantum potential $V_Q=-(1/2m) \nabla^2 \sqrt{\rho}/\sqrt{\rho}$, while $\rho$ can be viewed as the density of the fluid. Almost all particles are in the condensate, the effective pressure of which is computed to be
\begin{equation}
P = \left( \frac{2 \pi l}{m^3} \right) \epsilon^2 = K \epsilon^2. 
\end{equation}
In the  case, the scattering length, the mass of the particles and the scale $\Lambda$ are related via \cite{chavanis3,Fan}
\be
\frac{l}{m^3} = \frac{1}{32 \pi (F m)^2} = \frac{1}{32 \pi \Lambda^4},
\en
and therefore we obtain for the constant $K$ the final expression $K=1/(2 \Lambda)^4$.

%%%%%%%%%%%%%%%%%%%%%%%%%%%%%%%%%%%%%%%%%%%%%%%%%%%%%%%%%%%%%%%%%%%%%%%%%%%%%%%%

%\bibliographystyle{yahapj}

\end{document}